\documentclass[sigconf]{acmart}

\usepackage[utf8]{inputenc} 
\usepackage[T1]{fontenc}    
\usepackage{url}            
\usepackage{booktabs}       
\usepackage{nicefrac}       
\usepackage{microtype}      
\usepackage{xcolor}         
\usepackage{natbib}
\usepackage{cleveref}
\usepackage{subcaption}

\AtBeginDocument{%
  }

\copyrightyear{2024}
\acmYear{2024}
\setcopyright{rightsretained}
\acmConference[FAccT '24]{The 2024 ACM Conference on Fairness, Accountability, and Transparency}{June 3--6, 2024}{Rio de Janeiro, Brazil}
\acmBooktitle{The 2024 ACM Conference on Fairness, Accountability, and Transparency (FAccT '24), June 3--6, 2024, Rio de Janeiro, Brazil}\acmDOI{10.1145/3630106.3658948}
\acmISBN{979-8-4007-0450-5/24/06}

\begin{document}

\title{Visibility into AI Agents}

\author{Alan Chan}
\authornote{Correspondence to \texttt{alan.chan@mila.quebec}}
 \email{alan.chan@mila.quebec}
 \affiliation{%
  \institution{Centre for the Governance of AI}
  \city{Oxford}
\country{UK}
 }
\affiliation{%
  \institution{Mila (Quebec AI Institute)}
  \city{Montréal}
\country{Canada}
 }

 \author{Carson Ezell}
 \affiliation{%
  \institution{Harvard University}
  \city{Cambridge}
\country{USA}
}

 \author{Max Kaufmann}
 \affiliation{%
   \institution{Independent}
   \city{London}
   \country{UK}
 }

 \author{Kevin Wei}
 \affiliation{%
   \institution{Harvard Law School}
     \city{Cambridge}
\country{USA}
 }

 \author{Lewis Hammond}
 \affiliation{%
   \institution{University of Oxford}
   \city{Oxford}
\country{UK}
 }
 \affiliation{%
   \institution{Cooperative AI Foundation}
   \city{Oxford}
\country{UK}
 }

 \author{Herbie Bradley}
 \affiliation{%
   \institution{University of Cambridge}
   \city{Cambridge}
   \country{UK}
 }
 \author{Emma Bluemke}
 \affiliation{%
   \institution{Centre for the Governance of AI}
   \city{Oxford}
\country{UK}
 }
 \author{Nitarshan Rajkumar}
 \affiliation{%
   \institution{University of Cambridge}
   \city{Cambridge}
   \country{UK}
 }

\author{David Krueger}
 \affiliation{%
   \institution{University of Cambridge}
   \city{Cambridge}
   \country{UK}
 }
\author{Noam Kolt}
 \affiliation{%
   \institution{University of Toronto}
   \city{Toronto}
   \country{Canada}
 }
\author{Lennart Heim}
\authornote{Equal co-supervision.}
 \affiliation{%
   \institution{Centre for the Governance of AI}
   \city{Oxford}
\country{UK}
 }
\author{Markus Anderljung}
\authornotemark[2]
 \affiliation{%
   \institution{Centre for the Governance of AI}
   \city{Oxford}
\country{UK}
 }

\renewcommand{\shortauthors}{Chan et al.}

\begin{abstract}

Increased delegation of commercial, scientific, governmental, and personal activities to AI agents---systems capable of pursuing complex goals with limited supervision---may exacerbate existing societal risks and introduce new risks. Understanding and mitigating these risks involves critically evaluating existing governance structures, revising and adapting these structures where needed, and ensuring accountability of key stakeholders. Information about where, why, how, and by whom certain AI agents are used, which we refer to as \textbf{visibility}, is critical to these objectives. In this paper, we assess three categories of measures to increase visibility into AI agents: \textbf{agent identifiers}, \textbf{real-time monitoring}, and \textbf{activity logging}. For each, we outline potential implementations that vary in intrusiveness and informativeness. We analyze how the measures apply across a spectrum of centralized through decentralized deployment contexts, accounting for various actors in the supply chain including hardware and software service providers. Finally, we discuss the implications of our measures for privacy and concentration of power. Further work into understanding the measures and mitigating their negative impacts can help to build a foundation for the governance of AI agents.

\end{abstract}




\maketitle

\section{Introduction}\label{sec:intro}
Many AI developers are creating systems with greater autonomy, access to external tools or services, and an increased ability to reliably adapt, plan, and act open-endedly over long time-horizons to achieve goals \citep{chan_harms_2023,richards_auto-gpt_2023,ruan_identifying_2023,liu_agentbench_2023,sharkey_causal_2023,kenton_discovering_2023,sumers_cognitive_2023}. 
We will say that such systems possess relatively high degrees of \textbf{agency} and will refer to them as \textbf{(AI) agents} or \textbf{agentic systems} \citep{lieberman_autonomous_1997,chan_harms_2023,shavit_practices_2023}. 
Systems with relatively low degrees of agency are those that only aid human decision-making or produce outputs without acting in the world, such as image classifiers or text-to-image models. 
Examples of agents could include reinforcement learning systems \citep{adaptive_agent_team_human-timescale_2023,reed_generalist_2022} that interact extensively with the real world\footnote{Including the physical environment but also digital environments such as online platforms.} or more capable versions of language models with tool or service access that could, for example, plan and book a holiday or send an email on a user's behalf \citep{openai_chatgpt_2023,schick_toolformer_2023,ruan_identifying_2023,bran_chemcrow_2023}.

Current AI agents sometimes struggle to perform even simple tasks \citep{liu_agentbench_2023,kinniment_evaluating_2023,raji_fallacy_2022,valmeekam_can_2023,valmeekam_large_2023,mialon_gaia_2023}, but given increasing investments in AI research \citep{epoch_key_2023}, scaling laws \citep{barnett_direct_2023,kaplan_scaling_2020,hoffmann_training_2022}, pressures to develop autonomous capabilities for military use \citep{kott_intelligent_2018,horowitz_ai_2021,scharre_debunking_2021}, economic applications \citep{chan_harms_2023}, and scientific prestige \citep{ganguli_predictability_2022,chan_harms_2023}, we should not discount continued improvements in capabilities \citep{bowman_dangers_2022}.
Indeed, a core goal of the AI field since its inception has been to build agents \citep{sutton_reinforcement_2018,russell_artificial_2021}. 

As AI agents improve in capabilities, speed, and cost,\footnote{As an example of cost reduction, FLOP \citep{hobbhahn_trends_2023} or FLOP/s \citep{hobbhahn_trends_2022} per dollar could decrease at the same time as performance per FLOP increases \citep{erdil_algorithmic_2023}.} it may be easier and more competitive to delegate tasks currently done by humans to AI agents instead.
The development and deployment of agents has surged recently \citep{openai_introducing_2023,chase_langchain_2022,wu_autogen_2023,richards_auto-gpt_2023,shen_hugginggpt_2023} and could lead to the ubiquitous deployment of agents in commercial, scientific, governmental, and personal activities. Since such deployment may exacerbate existing risks and introduce new ones \citep{chan_harms_2023,shavit_practices_2023}, it is imperative to understand how to govern AI agents.

\subsection{Risks of AI Agents}\label{sec:intro-risks}
Rather than provide an exhaustive taxonomy of risks from AI agents,\footnote{See \citet{critch_tasra_2023,weidinger_sociotechnical_2023,shelby_sociotechnical_2023} for taxonomies of risks from AI systems and \citep{chan_harms_2023,shavit_practices_2023} for further discussion of risks from AI agents.} we highlight certain agent-specific risks. 
In comparison to risks from other AI systems, these risks focus on the potential for agents to 
remove humans from the loop \citep{korinek_preparing_2022,chan_harms_2023}. Without a human in the loop, agents may take multiple consequential actions in rapid succession and bring about significant impacts before a human notices. The ability to remove humans from the loop also means that an agent's task performance is less limited by the expertise of its user, compared to a situation where user must guide an AI system's actions or take actions themself.

\subsubsection{Malicious Use}
AI agents could be a large impact multiplier for individuals or coordinated groups who wish to cause harm \citep{shavit_practices_2023}. Existing AI systems have already assisted in malicious use, including voice cloning scams \citep{verma_they_2023} and fake news generation \citep{verma_rise_2023}. 
However, more capable AI agents could automate end-to-end pipelines for complex tasks that currently require substantial human expertise and time.
For untrained individuals, such agents could drastically increase the accessibility of engaging in severely harmful activities because no human in the loop would be required. 
For example, there is interest in building agents to execute scientific research, comprising autonomous planning and execution of scientific experiments \citep{bran_chemcrow_2023,boiko_emergent_2023}.
If such agents were to become as capable as human scientists, they might enable or accelerate the design and development of harmful tools (e.g., biological \citep{sandbrink_artificial_2023,soice_can_2023}, chemical \citep{boiko_emergent_2023,bran_chemcrow_2023,urbina_dual_2022}) for groups that currently lack the expertise for such production. Extremely persuasive AI agents may also enable and enhance influence campaigns \citep{karinshak_working_2023,bai_artificial_2024,hazell_large_2023}.

Understanding the extent to which agents will facilitate malicious use requires information about how they are used and how they interact with external systems \citep{weidinger_sociotechnical_2023}.
Moreover, when malicious users do cause harm with AI agents, regulatory enforcers will need measures to identify the users and hold them accountable.

\subsubsection{Overreliance and Disempowerment}
Overreliance on AI agents to automate complex, high-stakes tasks could lead to severe consequences. Humans can already rely on certain automated systems more than is warranted \citep{cummings_automation_2004, endsley_out---loop_1995, endsley_level_1999}. More capable agents may enable automation of an increasing array of complex and useful tasks. Users---including both individuals and institutions---may rely on agents even in high-stakes situations, such as interfacing with the financial or legal systems, because human alternatives (e.g., hiring a lawyer) may become relatively slower and more expensive. At the same time, these agents may malfunction for a variety of reasons, including design flaws \citep{raji_fallacy_2022,ngo_alignment_2022,zwetsloot_thinking_2019} or adversarial attack \citep{willison_prompt_2023,bailey_image_2023,zou_universal_2023}. Malfunction may not be immediately apparent, especially if users lack the requisite expertise or domain knowledge. Stopping the agent may be difficult if doing so would lead to cascading failures or a competitive disadvantage for the user \citep{shavit_practices_2023}.
More broadly, profit and efficiency motives may lead to collective dependence on agents for essential societal functions, such as the provision of government services \citep{de_la_garza_states_2020,zilka_transparency_2022} or the operation of essential infrastructure \citep{degrave_magnetic_2022,biagioni_powergridworld_2022}.   
Companies providing access to AI agents would hold substantial power \citep{burrell_society_2021}, while malfunction of those agents could have societal-scale impacts. At minimum, societies require information about the extent of reliance upon AI agents and whether such reliance is justified.

\subsubsection{Delayed and Diffuse Impacts}
Potential negative impacts of AI agents may be delayed and diffuse.\footnote{Roughly speaking, we consider an impact to be diffuse if it is difficult to observe and most apparent in aggregate across many individual cases.} 
Delayed and diffuse impacts may be difficult to manage because they may require sustained attention over long periods of time even to notice. 
Impacts of agents may be delayed if users give agents long-horizon goals, while diffuseness of impact may come from the widespread deployment of agents to automate complex processes. 
Consider an agent given the goal of continually finding and hiring job candidates who will most contribute to the company over the long-term. 
This agent may screen résumés \citep{gan_application_2024}, perform interviews, make the final hiring decision, and analyze the performance of hires. 
Given the time horizon over which the agent is acting and its influence over the company, any potential problems like algorithmic bias \citep{raghavan_mitigating_2020} could be hard to identify and become deeply entrenched. 
The most severe consequences of such problems may only be apparent when looking at how companies in aggregate use AI agents for hiring. 
AI agents could also subtly benefit their developers, akin to the self-preferencing behaviour of large-scale digital platforms \cite{kittaka_self-preferencing_2023}. 
Moreover, agents that mediate or even substitute for human communication \citep{noauthor_characterai_2023,melley_judges_2024} could have diffuse and delayed psychological and social impacts \citep{jakesch_co-writing_2023,bai_artificial_2024,karinshak_working_2023}, 
analogous to certain effects of social media \citep{bond_61-million-person_2012,sheffield_is_2021,milli_engagement_2023}. 
The deployment of agents may also induce changes in market structures or workforce impacts from job displacement \citep{acemoglu_artificial_2018,acemoglu_automation_2019}. Identifying delayed and diffuse impacts may require long-term tracking of the extent and nature of AI agent usage across a wide range of application areas.

\subsubsection{Multi-Agent Risks}
Interactions and dependencies between many deployed agents could lead to risks not present at the level of a single system \citep{green_emergence_2023,park_generative_2023,schelling_micromotives_1978,hammond_multi-agent_2024}. 
Agents could enter into destabilising feedback loops, such as those between automated trading algorithms in the 2010 flash crash \citep{ctfc_preliminary_2010}. 
Agents partially built upon the same components---such as a particular foundation model---could have common vulnerabilities and failure modes \citep{bommasani_picking_2022,cohen_complex_2010}; widespread deployment of such agents could risk large-scale systemic harms. 
More generally, there may be unpredictable behavioural changes that are characteristic of complex systems \citep{siegenfeld_introduction_2020,cohen_complex_2010,schelling_micromotives_1978}. 
Competitive pressures and selection effects could lead to the development of agents that act in more anti-social ways \citep{xu_language_2023,hendrycks_natural_2023,cassidy_how_2009,ely_natural_2023}. These potential issues motivate understanding not just individual agents, but also interactions within groups of agents.

\subsubsection{Sub-Agents}
Agents could instantiate more agents to accomplish (components of) a task, which may magnify several of the risks discussed so far.
It may be advantageous for an agent to create potentially specialized and more efficient \textbf{sub-agents}, especially if doing so is cheap and fast.
For example, an agent could call copies of itself through an API, or itself train, fine-tune, or otherwise program another agent.
Sub-agents could be problematic because they introduce additional points of failure; each sub-agent may itself malfunction, be vulnerable to attack, or otherwise operate in a way contrary to the user's intentions. 
Stopping an agent from causing further harm might involve intervening not only on the agent, but also on any relevant sub-agents \citep{carey_human_2023,soares_corrigibility_2015}. Yet, this process may be difficult because we lack methods for determining when an agent has created a sub-agent. 
Information about the extent of sub-agent creation and operation can enable a better understanding of the significance of these risks.

\subsection{The Case for Visibility into AI Agents}\label{intro:visibility}

Addressing the risks of AI agents requires \textbf{visibility}:\footnote{We use \textit{visibility} rather than \textit{transparency} as we believe the former to be somewhat more common in a regulatory context. Both terms are distinguished from \textit{explainability}, which refers to whether one can understand why an AI system generated a particular output \citep{miller_explanation_2019}} information about where, why, how, and by whom AI agents are used. 
Visibility would help to evaluate existing governance structures, revise and adapt these structures where needed, and ensure accountability of key stakeholders. 
Regulatory oversight bodies which monitor and enforce rules on the activities of human agents and certain automated programs (e.g., trading algorithms)
\citep{laffont_theory_2002,holmstrom_moral_1979,holmstrom_multitask_1991,ball_workplace_2010,jensen_theory_1976,birkinshaw_freedom_2006,minkkinen_continuous_2022,van_loo_regulatory_2019,coglianese_seeking_2004,board_of_governors_of_the_federal_reserve_system_proactive_2021, cftc_cftc_2023,bielicki_monitoring_2020,guarnieri_surveillance_2021} may require additional information to understand and address harms from AI agents. 
For example, if agents are able to employ novel strategies for collusion \citep{dorner_algorithmic_2021} when carrying out economic activities, new rules and updates to investigative authority may be necessary. 
Furthermore, AI agents may simultaneously provide services traditionally regulated by different agencies, such as both financial and legal services.
The same agent developer or deployer may thus exercise power across diverse and usually independent domains of regulation, creating additional concerns related to market consolidation and conflicts of interest.

Visibility measures also play a central role in addressing problems that arise when humans delegate to other humans or institutions \citep{laffont_theory_2002}. 
The precise purpose of such regimes varies, and can include reducing information asymmetries, shaping incentive structures, and triggering enforcement actions \citep{holmstrom_moral_1979,holmstrom_multitask_1991}. For example, employers often monitor the conduct and performance of employees through ongoing supervision and periodic performance reviews \citep{ball_workplace_2010}. In corporate governance, shareholders monitor management through a range of institutional mechanisms, including financial audits, shareholder meetings, and company reports, buttressed by legally binding fiduciary duties (in the case of directors) and the ability (in some cases) to dismiss management if they fail to act in the collective interest of shareholders \citep{jensen_theory_1976}.
Comparable mechanisms exist to support citizens in monitoring the activities of government bodies and public officials. 
These mechanisms include maintaining records of government decisions, facilitating information access through freedom of information requests, and commissioning detailed public reports into government activities \citep{birkinshaw_freedom_2006}, a combination of which may ultimately inform citizens’ electoral choices. 
Although visibility measures can be costly \citep{fama_agency_1983} and raise privacy concerns, 
they remain a necessary feature of frameworks for shaping the incentives of, and governing, agents. 

We emphasize visibility into \textit{deployed} AI agents because the scope and severity of potential impacts may not be apparent during development. By \textbf{deployed}, we mean agents that are in use, whether the agent is available to the general public, select customers, or only for internal use within the organization that develops it. Visibility into the last case may be particularly important if organizations that carry out crucial societal functions, such as banks or cloud compute providers, develop and deploy their own AI agents. 
We focus on deployment because pre-deployment testing \citep{shevlane_model_2023,anderljung_frontier_2023} does not account for how users or deployers may exacerbate risks \citep{weidinger_sociotechnical_2023} through fine-tuning \citep{davidson_ai_2023}, connecting to external tools or services \citep{openai_chatgpt_2023,sharkey_causal_2023,openai_introducing_2024,openai_introducing_2023}, or structuring calls to the system so as to better enable it to pursue goals \citep{richards_auto-gpt_2023,wu_autogen_2023,sharkey_causal_2023,wei_chain--thought_2023}. 
Even instances of agents that come from the same underlying system can access different tools and can be conditioned to behave differently based on prompts.

\subsection{Contributions}

In this paper, we assess three categories of measures to increase visibility into AI agents: agent identifiers, real-time monitoring, and activity logs.
For each, we outline potential implementations that vary in intrusiveness of data collection and informativeness of the data.
We analyze how our measures apply %
across a spectrum of centralized through decentralized deployment contexts, accounting for various actors in the supply chain including hardware and software service providers. Finally, we discuss the implications of our measures for privacy and concentration of power. Rather than advocating for immediate implementation of these measures, we emphasize the need for further understanding them and how to mitigate their negative impacts. 

The measures extend existing work in deployment visibility to better account for the risks of AI agents. Agent identifiers, which indicate whether and which AI agents are involved in interactions, generalize watermarks \citep{liu_survey_2024,wang_data_2021} because they apply to all of an agent's outputs, including the use of external tools and services, not just text, image, or audio outputs. This generalization is crucial for improving visibility if agents increasingly substitute for human actions. For real-time monitoring and activity logging, we assess practices that extend existing schemes \citep{henzinger_runtime_2023,shergadwala_human-centric_2022,javadi_monitoring_2020,javadi_monitoring_2021,minkkinen_continuous_2022} so as to better track complex interactions between multiple agents \citep{schelling_micromotives_1978,critch_tasra_2023}, an agent's interaction with external tools or services, and delayed and diffuse effects of an agent's actions.

\begin{figure*}
    \centering
    \includegraphics[width=\textwidth]{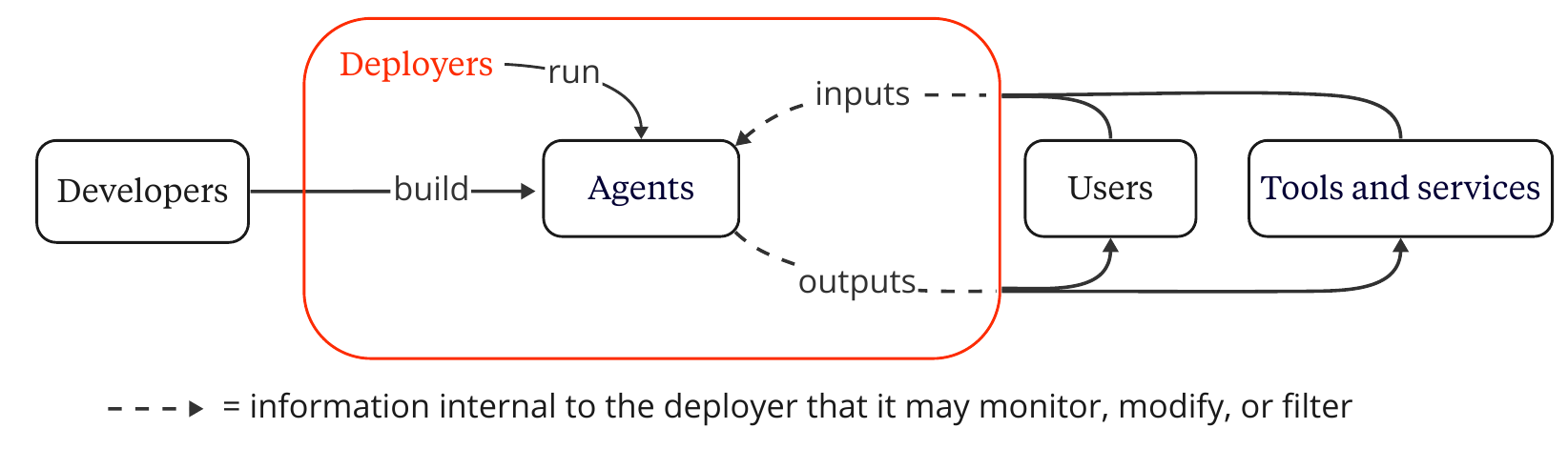}
    \caption{We illustrate how our main terms in \Cref{sec:definitions} interact with each other. \textcolor{red}{Deployers} are in red and encompass the \textbf{agents} box to denote the fact that our paper focuses on agents that are run by deployers and served to users. Developers build agents (or an underlying system) and \textcolor{red}{deployers} serve instances of agents to users. Since deployers run agents, the inputs and outputs of agents are by default visible to the deployer, which facilitates the measures that we discuss in \Cref{sec:measures}.} 
    \label{fig:defs}
\end{figure*}

\section{Definitions}\label{sec:definitions}
Besides the definitions here, we also define each term when we use it for the first time in the main body. We illustrate the most common terms in \Cref{fig:defs}.

\textbf{Agency} is the degree to which an AI system 
acts directly in the world to achieve long-horizon goals, with little human intervention or specification of how to do so. 
An \textbf{(AI) agent} is a system with a \textit{relatively high} degree of agency; we consider systems that mainly predict without acting in the world, such as image classifiers, to have relatively low degrees of agency. 
Examples of agents include reinforcement learning systems \citep{adaptive_agent_team_human-timescale_2023,reed_generalist_2022} that interact extensively with the real world\footnote{Including the physical environment but also digital environments such as online platforms.} or more capable versions of language models with tool access \citep{openai_chatgpt_2023,schick_toolformer_2023,ruan_identifying_2023,bran_chemcrow_2023}. We do not consider existing foundation models themselves to be agents. 
Our definitions compress the characterization of agency in \citet{chan_harms_2023}, which points to four axes: the degree to which the system's behaviour is specified, the degree to which the system's behaviour is goal-directed, the degree to which the system has a direct impact in the world, and the degree to which the system can achieve goals over long time-horizons. For the purposes of this paper, we use agent and \textbf{agentic system} interchangeably. 

\textbf{Scaffolding} is any method that structures the calls to an AI system so as to facilitate the pursuit of goals \citep{richards_auto-gpt_2023,wu_autogen_2023,chase_langchain_2022}. Scaffolding may include additional prompts, memory systems, access to external tools, and planning mechanisms \citep{wang_survey_2023}. For example, AutoGPT \citep{richards_auto-gpt_2023} has a language model accept a high-level goal and sequentially produce \textit{(reasoning, plan, criticism of the plan, action)} tuples so as to achieve the goal. Scaffolding can make an AI system, such as a foundation model, more agentic. 

The term \textbf{developer(s)} refers to the actor(s) involved in the construction of an AI system. While the developers of a system include those who trained the underlying machine-learning model, developers could also include those who build other components of the complete system, such as the scaffolding \citep{richards_auto-gpt_2023,wu_autogen_2023}. 

The \textbf{user} is the human individual or group that interacts with and provides instructions to an AI system.  

The \textbf{deployer} is the entity that operates an AI system and serves it to users. The deployer may not be the same as the developer(s). For example, Microsoft deploys OpenAI's systems into its products \citep{warren_microsofts_2024}, but did not develop GPT-4.  
A deployer may provide access to an agent in one of two ways. First, the deployer may serve a foundation model which users may combine with other components to make a more\footnote{The system may not be completely autonomous since user approval may still be required for certain actions.} agentic system. For example, users may use the scaffolding framework AutoGPT \citep{richards_auto-gpt_2023} to chain calls to GPT-4 \citep{openai_gpt-4_2023} and provide the model with tool access. 
Indeed, popular scaffolding frameworks depend upon an underlying foundation model \citep{wu_autogen_2023,richards_auto-gpt_2023}. 
Second, the deployer may provide an agent or may furnish ways for users to make a provided system more agentic. For instance, OpenAI lets users build and share custom agents \citep{openai_introducing_2024,openai_introducing_2023} augmented with a variety of tools, including browsing, using Google Drive apps, and coding \citep{openai_chatgpt_2023}.

The \textbf{compute provider} is responsible for supplying and maintaining the hardware infrastructure on which an AI system operates. The compute provider could also be the same as the deployer or the developer if either runs its own infrastructure. Compute providers could be important partners for overseeing large-scale deployments of agents that are not run by deployers, which we discuss further in \Cref{sec:non-deployer-agents}. 

A \textbf{tool} or \textbf{service} refers to an external system or platform with which an AI agent interacts to perform its tasks. For instance, a flight booking website where the AI agent executes transactions, such as purchasing plane tickets on behalf of the user, would be a service. The \textbf{provider} of the tool or service is responsible for maintaining the system or platform. We will often use \textit{tool} and \textit{service} interchangeably. Agents often interact with tools or services through dedicated \textbf{APIs}, which are interfaces and protocols specifically structured for agents, rather than human users.

The \textbf{outputs} are the results or responses that an AI system generates. Some types of outputs include images, text, or actions (e.g., calling a tool). 
While the deployer generally has access to all the outputs, the tool or service provider's knowledge is limited to outputs relevant to their specific service (e.g., the results of an API call).

\textbf{Inputs} are the data that the AI agent receives from a user, a tool or service, another agent, or any other party, which inform its actions or responses. 
The deployer in principle has access to inputs by virtue of running the system, but may choose not to collect or store such information out of respect for user privacy.

\section{Measures to Improve Visibility}\label{sec:measures}
\begin{figure*}
    \centering
     \begin{subfigure}[b]{\textwidth}
         \centering
         \includegraphics[width=0.8\textwidth]{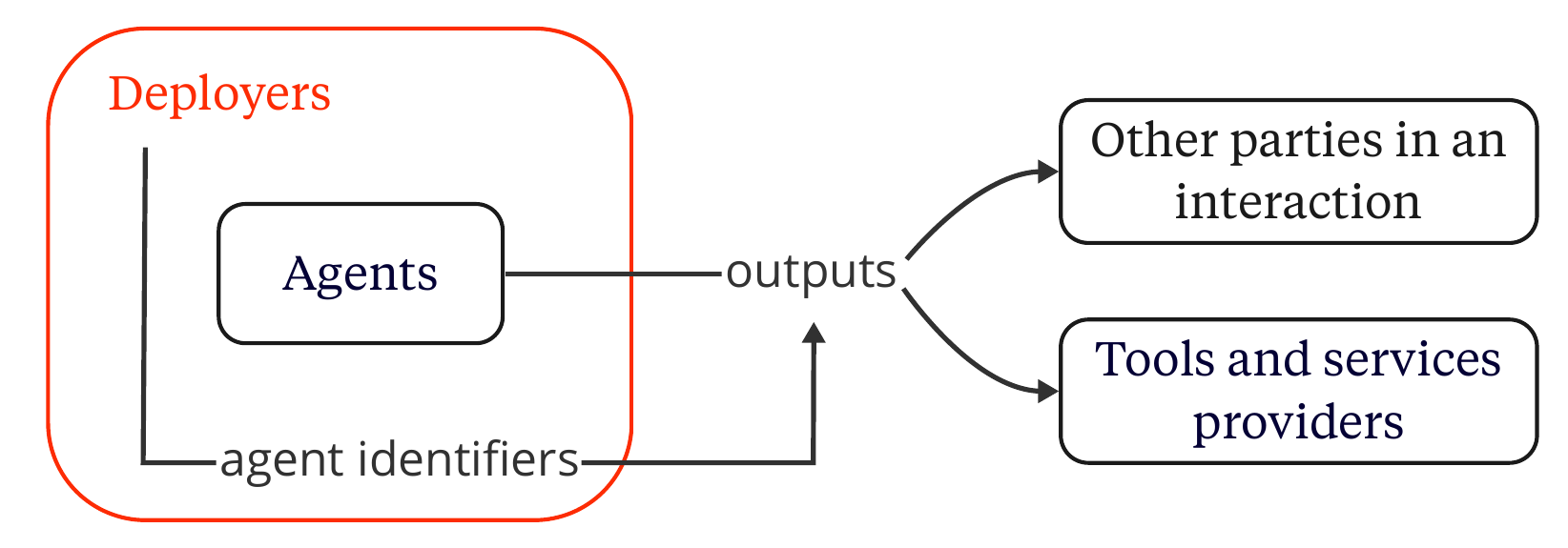}
         \caption{An \textbf{agent identifier} indicates to certain actors whether an AI agent is involved in an interaction. Developers (not shown) and \textcolor{red}{deployers} cooperate to implement agent identifiers, which the latter adds onto outputs. Here, we illustrate an agent identifier that informs other parties in a given interaction with an agent, as well as tools and services providers. If these actors know that they are interacting with an agent, they may wish to verify certain properties such as the security or robustness of the agent. See \Cref{sec:agent–identifiers} for further discussion.}
         \label{fig:identifiers}
     \end{subfigure}
     
     \vspace{1em}
     
     \begin{subfigure}[b]{\textwidth}
         \centering
         \includegraphics[width=\textwidth]{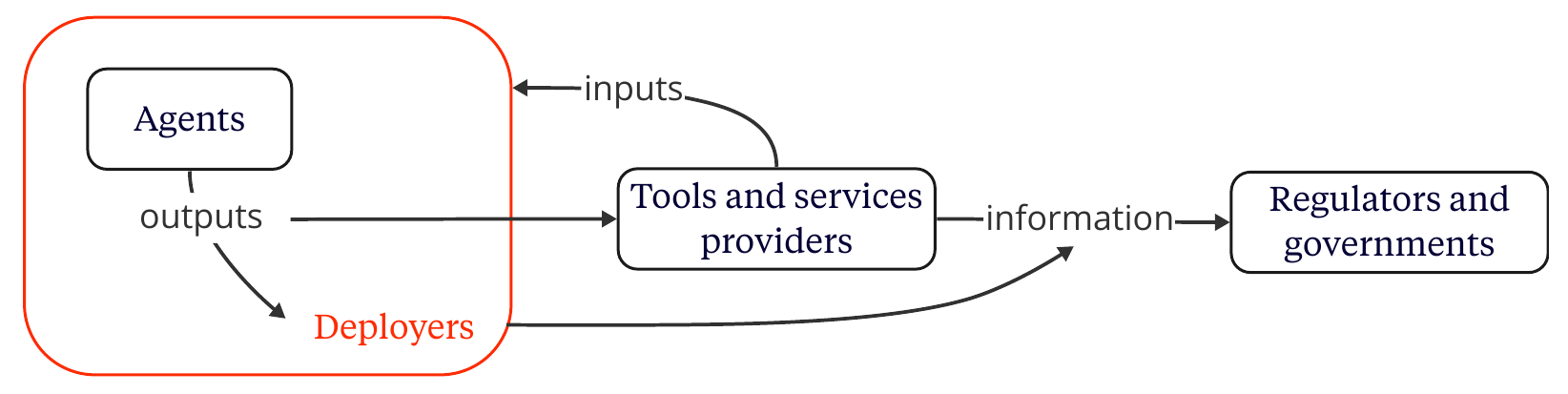}
         \caption{An agent's inputs and outputs are visible to the \textcolor{red}{deployer}. Inputs come from tool and service providers and users (not shown). Certain outputs, such as requests to external tools and services, are also visible to tools and services providers. These actors can monitor and filter the actions in real-time (\Cref{sec:real-time}) or keep logs (\Cref{sec:logs}) for post-incident attribution or forensics. Insights gained from real-time monitoring or from logs can inform regulators and governments.}
         \label{fig:logging_monitoring}
     \end{subfigure}
     \caption{We illustrate the flow of information for our measures in \Cref{sec:measures}.}
     \label{fig:measures}
\end{figure*}

We propose three complementary categories of measures to improve visibility into AI agents. \textbf{Agent identifiers} indicate whether and which AI agents are involved in a given interaction, such as watermarks or IDs that distinguish agents in their requests to service providers. \textbf{Real-time monitoring} involves real-time analysis of an agent's activity, allowing deployers and/or service or tool providers to flag and intervene on problematic behaviour as it is occurring. 
\textbf{Activity logs} held by deployers and tool or service providers record certain inputs and outputs of an agent, such as interacting with external services or other agents, to facilitate post-incident attribution and forensics. See \Cref{fig:measures} for an overview of the information flows for each measure. 

Each category contains measures that vary in intrusiveness of data collection and informativeness. More comprehensive information collection may be justified for agents deemed to be high-risk, potentially based on the results of evaluations \citep{kinniment_evaluating_2023,anderljung_frontier_2023,shevlane_model_2023,ruan_identifying_2023,mialon_gaia_2023,rismani_plane_2023} or deployment in high-risk domains \citep{noauthor_proposal_2021,koessler_risk_2023,critch_tasra_2023}. For example, it may be desirable to subject agents involved in financial trading to monitoring requirements at least as strict as those for human traders \citep{division_of_trading_and_markets_guide_2008}. 
Yet, more comprehensive data collection may have serious privacy risks, which we discuss in \Cref{sec:measures-privacy}. Our goal is to provide an array of options, rather than an answer to these trade-offs and the extent to which visibility measures should be mandated. Finally, while we discuss potential implementations of the measures, more research is required to understand their feasibility and implications. 

We focus in this section on agents run by \textbf{(agent) deployers}---entities that deploy agents, or important subcomponents like a foundation model, as a service to users. We include foundation models because many frameworks for constructing agents use a foundation model as the central component \citep{wu_autogen_2023,richards_auto-gpt_2023,openai_chatgpt_2023}. 
While deployers are unlikely to account for all agent activity, they likely constitute a substantial fraction because the most capable foundation models are only available through deployers \citep{openai_gpt-4_2023,gemini_team_gemini_2023,bai_constitutional_2022},\footnote{Currently, these deployers also happen to be developers.} 
and using a deployer may be more convenient than running a system oneself. Moreover, since deployers can already see the inputs and outputs of deployed systems, they can attach agent identifiers to outputs, perform real-time monitoring, and collect activity logs. In \Cref{sec:non-deployer-agents}, we analyze how to extend our measures to decentralized deployments of agents.

\subsection{Agent Identifiers}\label{sec:agent–identifiers}
An agent identifier indicates whether and which AI agents are involved in interactions. 
Agent identifiers are attached to select outputs, are visible to certain actors, and may include additional information about the agent. 

The ability to identify agents could be useful to several actors. A regulator could require AI agents to identify themselves as non-humans during interaction with humans \citep{frazier_right_2023,baker_eu_2023}, similar to bot disclosure laws \citep{diresta_new_2019}.\footnote{Note that not all non-human activities may come from AI agents. For example, consider currently automated trading activity or ads auctions.} Members of the general public may wish to know whether they have interacted with AI agents. Summary statistics based on agent identifiers could inform governments and the general public about the extent to which agents operate in high-risk settings \citep{noauthor_proposal_2021}. Identifiers for when agents send requests to tools or services providers could help to identify significant actions, such as when agents transfer sensitive information. A service provider may even reject a request absent certain guarantees attached to the identifier, such as those related to the security of the agent. Unique identifiers for each AI agent could facilitate accountability by linking an action to an AI agent and its user, developer(s), and deployer.

\subsubsection{Types of Agent Identifiers}
We consider three key design decisions for an agent identifier:

\textbf{Which outputs contain the identifier?}
Decisions to attach an identifier may consider both the format and content of the output. By format, we mean whether the agent outputs data such as images, text, audio, or API requests to a service provider. An agent identifier's specific implementation depends upon the output format. For example, identifiers for image outputs could be watermarks \citep{liu_survey_2024,wang_data_2021}, while an identifier for an API request could be a simple header, similar to headers in HTTP requests. The difficulty of implementing identifiers varies based on the format of the output: for instance, adversarial users may easily remove watermarks \citep{zhang_watermarks_2023}. Regarding content, identification may be especially important for significant outputs. 
Certain outputs, such as purchases made on behalf of the user, may merit identifiers by virtue of the task the agent is accomplishing. Other outputs may only be significant beyond a certain threshold, such as requests for compute resources that exceed a certain amount. 

\textbf{Which actors can see the identifier?}
An identifier may need to be visible to different actors. 
For example, in the context of a financial transaction, an agent identifier could be visible to any combination of the bank, the other party in the transaction, or the service provider for the bank API (which could be the bank itself). Some actors may need agent identifiers to fulfil their existing duties, such as e-commerce websites which must authenticate users and safeguard customer payment information 
\citep{pci_security_standars_council_payment_2022}. Furthermore, facilitating the identification of multi-agent risks may require that agent identifiers be visible to other agents.

\textbf{How specific is the identifier to a particular agent?}
An identifier could point to a particular agent, or simply denote that \textit{some} agent was involved in the interaction. 
The former could facilitate incident reporting and investigation. To implement unique identifiers for each deployed instance of an agent, cryptographic methods---such as those used in software attestation---may be needed to assure the agent's provenance. In \Cref{sec:agent-card}, we discuss additional, useful information that could be attached to an identifier.

\subsubsection{Attaching Additional Information to Agent Identifiers}\label{sec:agent-card}
Additional information attached to an agent identifier may be of further use. Additional information could be specific to the instance of the agent deployed to the user, or could pertain to the \textbf{underlying system} used during agent development. 
Information about the former could include the goals the user has given its agent, while information about the latter could include the results of evaluations performed on the underlying system. 
We refer to the set of such additional information as an \textbf{agent card}, drawing inspiration from previous work on documenting AI systems \citep{mitchell_model_2019,gebru_datasheets_2021,gilbert_reward_2023,bommasani_ecosystem_2023}. 
In \Cref{app:agent-card} we provide a more comprehensive list of what could be included on an agent card, but in this section we discuss three particularly important types of information that could be included. 

\textbf{The underlying system.}
This information could include the results of evaluations \citep{shevlane_model_2023}; previous incidents; the dependencies involved in the system's construction \citep{mitchell_model_2019,gebru_datasheets_2021,bommasani_ecosystem_2023}; or training methods and data used. Such information could inform the decisions of actors that interact with agents. For example, tool providers may reject requests from agents that do not meet certain security standards \citep{bhatt_purple_2023}.

\textbf{The specific instance of the agent.}
This information could consist of how the agent was deployed (e.g., by its user directly or by another agent); a list of external tools or services that the agent can access (e.g., applications or software, any physical tools); the scaffolding framework (e.g. memory or planning mechanisms); the intended scope, permissions, and goal(s) of the agent \citep{gilbert_reward_2023}; or the intended sector of deployment (e.g., finance). This type of information could be useful for regulators to develop an overall understanding of the extent and significance of agent use. 

\textbf{The actors involved in agent development and deployment.}
These actors could include the user, the compute provider, developers, and other humans with whom the agent has interacted. Authentication of the agent's user may be useful when individuals are providing sensitive information to agents (e.g. agents used by banks to assist customers). Furthermore, this information helps to clarify \textit{whom} to hold accountable in case an agent causes harm during its operation 
\citep{nissenbaum_accountability_1996,cooper_accountability_2022}. 
A regulator could use identifying information associated to each of the actors, ideally in synergy with data from \Cref{sec:real-time,sec:logs}, to help investigate and address problems. 
For example, a regulator may receive reports from third parties of an agent engaging in financial scams. 
If the transactions of the agent contain agent identifier, a regulator could match the identifier to the corresponding agent card and ask the deployer of the agent to investigate further or shut it down. 

A more comprehensive agent card may be more useful, but may also be more intrusive to collect and require more substantial resources and expertise for analysis. Comprehensiveness of information should likely scale with the risk of the activity in question. 

\subsection{Real-Time Monitoring}\label{sec:real-time}
Real-time monitoring involves oversight of agent activity in real-time, with the goal of flagging and potentially filtering problematic behaviour. Such monitoring is likely to be automated so as to be able to cope with the speed and scale of agent activities \citep{shavit_practices_2023,greenblatt_ai_2024}; see \citet{naihin_testing_2023} for a preliminary implementation of a real-time monitoring system for agents. Crucially, real-time monitoring does not require the collection or storage of activity logs. As an analogy, consider the difference between moderation filters and chat logs in the context of a text messaging application---moderation filters might instantly delete harmful content that trips a filter, but this measure is separate from keeping logs for subsequent analysis. This difference can be critical for users who are concerned about the privacy implications of a deployer keeping logs of user inputs, such as business customers who need to consider the security of their corporate IP. Indeed, some cloud providers already offer no-logging provisions\footnote{For feasibility, sometimes logs must be kept for a short amount of time. The no-logging provision of \citet{google_bard_2023} keeps logs for 72 hours.} for their language model deployments to some customers, subject to real-time monitoring for abuse \citep{chrishmsft_data_2023,mrbullwinkle_azure_2023,google_bard_2023}.

Once behaviour is flagged, a deployer can intervene on the agent and potentially inform relevant authorities. A user could be given the opportunity to request human review of any action taken to impede its agent. 
We leave details about the implementation challenges of automation and human review to other work \citep{sullivan_explanation_2022,green_flaws_2022,venkatasubramanian_philosophical_2020,karimi_algorithmic_2021}.

Since real-time monitoring requires an immediate decision to flag a behaviour, deployers would only be able to find problems which they can identify in real-time. Thus, real-time monitoring may be most suitable for identifying violations of clear rules such as restrictions on allowed activities of agents, limits on tools and permissions made available to agents, leakage of sensitive information, or usage caps for particular users or agents. Behaviors can also be flagged for anomalies.

Flagging behaviors might also involve combining information from multiple agents because an individual agent's behaviour may only be problematic in the context of other agents. 
For instance, collusion requires a group of agents, not just a single agent, to coordinate to set high prices \citep{dorner_algorithmic_2021}.
Similarly, a single agent rapidly moving resources (e.g., data or financial investments) between entities may not be an issue in isolation, but vast numbers of agents engaging in the same behaviour might overload a given system (e.g., server outages or stock crashes). 
A real-time monitoring system could use agent identifiers to track which agents were involved in interactions together. This system could raise flags if indicators of the collective activity of these agents pass certain thresholds. For example, an indicator for algorithmic collusion could be correlation in price movements. 

Real-time monitoring could also flag when an agent exerts significant influence in its environment \citep{pan_rewards_2023}. Such activities could include economic transactions above a certain amount, usage of substantial compute resources, running a large number of sub-agents, or simultaneous interfacing with a large number of physical systems. While the acquisition of significant influence may not itself cause harm, the impact of an agent's failure would be magnified. A potential response to such a flagged activity could be pausing it until the deployer can verify explicit user approval, just as banks monitor transactions and temporarily suspend accounts if they suspect fraud.

\subsection{Activity Logs}\label{sec:logs}
Activity logs are records of certain inputs and outputs of an agent. Inputs and outputs could be from both users or tools and services, as we illustrate in \Cref{fig:defs}. Records could also report changes in the state of the agent, such as updates to the weights. Logs can include varying levels of detail and may only record actions above a certain degree of significance, depending on the behaviour one is trying to track, privacy considerations, and storage requirements. Frameworks for building agents already implement logging of certain actions \citep{chase_langchain_2022,richards_auto-gpt_2023,wu_autogen_2023}. 

Activity logs can inform an understanding of the agent's impacts as well as allow one to tie behaviour to specific user choices, facilitating post-incident attribution and forensics \citep{obrien_deployment_2023}. During audits and incident investigation, investigators may need detailed information to trace the source and causes of harms arising from AI agents' actions \citep{mokander_auditing_2023,raji_outsider_2022}. Researchers may benefit from detailed logs to improve our understanding of agents and to develop better deployment controls. 
Analysis of logs may also identify novel behaviours to improve real-time monitoring.

\subsubsection{Agent-Specific Information}
While activity logs are useful for understanding risks from AI systems in general \citep{javadi_monitoring_2020,javadi_monitoring_2021} and are common practice in many domains \citep{iso_isoiec_2022,noauthor_12_2022,noauthor_14_2023,securities_and_exchange_commission_17_2022,securities_and_exchange_commission_17_2022-1}, they can be especially helpful for addressing the risks we identified in \Cref{sec:intro-risks}. Different categories of logs could exist, such as for tool use, internal reasoning \citep{wei_chain--thought_2023}, self-critique \citep{saunders_self-critiquing_2022}, memory management, etc, each of which may be useful to different actors.

Understanding the impacts of agents involves identifying which scaffolding, tools, and services they use. For example, an agent with external memory can plan and act over longer horizons. Logs of how an agent manages its memory may be helpful for a user in trying to understand the agent. 
More generally, a log could explicitly label when an agent has accessed an external tool and the tool's output. Both the deployer and the tool provider could engage in such logging: deployer logs may be more useful for understanding how the tool affects the overall behaviour of the agent, while tool provider logs may provide more insight into the impacts of the tool itself. Indeed, tool providers may have strong incentives to engage in such logging themselves: for example, tool providers can study logs to update APIs or user interfaces to prevent abuse. Tool providers can also decide to restrict services to certain AI agents with identifiers (and potentially other attached certifications), as discussed in \Cref{sec:agent–identifiers}.

Identifying delayed and diffuse impacts may require logs to be retained for extended periods of time. 
Details about the persistence of the agent could be included in the logs, such as its running time, whether it is writing to and accessing external memory, or the amount of compute used so far to run the agent. These details could inform interventions, such as limiting the lifetimes of certain agents. Yet, significant impacts may arise after the lifetime of an individual agent. The impact of the original agent could be delayed, or a user could run another agent for the same purposes, potentially even with the same inputs and memory as the original agent. Accounting for these possibilities means that logs may have to persist for a significant amount of time after the lifetime of the corresponding agent. Furthermore, logs for different agents may have to be combined if one agent can be viewed as a continuation of another. 

Combining information from multiple logs may also help to understand sub-agent and multi-agent dynamics. For example, agent logs could be used to build models of how a particular malfunction might propagate through a network of agents or identifying undesirable forms of communication between agents \citep{roger_preventing_2023}.

\subsubsection{Logging at Different Levels of Detail}
A key design decision is the level of detail at which to record the agent's actions. Less detailed logging may only record high-level summaries of agent's behaviour or certain samples thereof. 
At the finest level, a regulator may require a deployer to record in detail all of an agent's behavior, especially if an agent is operating in a high-risk environment. More detailed logging is more useful, but may impose more significant costs on the deployer, require more resources and expertise for analysis, and pose more significant privacy concerns.

\subsection{Risks}\label{sec:measures-privacy}
Privacy considerations may conflict with obtaining detailed information about agent activity. Language model deployers are increasingly offering customers, particularly business customers, privacy assurances around data collection and use. Measures to reassure customers about confidentiality include:
\begin{itemize}
    \item Language model APIs with no logging of inputs or outputs, and the ability to turn off safety filters and moderation classifiers \cite{chrishmsft_data_2023}.
    \item Guarantees that customer data, including system outputs, will not be used to train any AI system and will be kept in a customer's cloud instance \cite{google_cloud_generative_2023}.
    \item The ability to delete logs kept by the provider after a certain amount of time \cite{chrishmsft_data_2023}.
\end{itemize}
Additionally, existing data protection laws, such as GDPR, impose further restrictions. Agent cards may contain identifying information about users. Agent logs may be considered personal data, such as when agents are given access to a filesystem containing personally identifiable information \cite{finck_they_2020}. 

In general, if agents substitute for humans in a wide variety of activities, information about those agents might be tantamount to information about the users of those agents. Indeed, agent activities may be easier to monitor than human activities because deployers are a central intermediary. Governments or deployers may thus abuse their power to carry out excessive or unjustified surveillance of personal activities \citep{goodwin_cooperation_2018,greenwald_nsa_2013}. These considerations justify limiting data collection in accordance with the risk of the agent's activities or domain of deployment. Another potential mitigation may be decentralized data custody schemes or data trusts \citep{delacroix_bottom-up_2019,delacroix_democratising_2020} whereby users or accountable representatives would make decisions about data usage. 

Modulating the degree of access to collected information can also help to mitigate privacy concerns. 
Access can vary with respect to \textbf{granularity}, the amount of detail contained in the records, and \textbf{quantity}, the number of records that a party is allowed to access. With respect to granularity, information can be aggregated, de-identified, or identifiable. Aggregated information involves summary statistics but not individual records or logs; differentially private \citep{bluemke_exploring_2023} computations of summary statistics may help to preserve the privacy of individual records. Records and logs can be de-identified with respect to individual users or identifiable. With respect to quantity, a party can have full access to all records, access based on approved search queries or filters, or access upon-request to pre-specified records to which they must provide a compelling reason for access. 

The granularity and quantity of access should be the minimum necessary for the accessing party to achieve its (legitimate) objectives. When investigations pertain to specific users, identifiable information could be made available upon request given a showing of compelling need and/or after approval from a third-party adjudicator. Regulators may need logs containing identifiable information in some cases, such as oversight of certain high-risk or high-volume activities. For example, 
for traders transacting above a specified threshold, CFTC collects identifiable personal information to enable aggregation of data across different accounts and brokers \citep{cftc_large_2023}.

\section{Decentralized Deployments}\label{sec:non-deployer-agents}
Some deployments of agents may occur in a decentralized way and bypass deployers. 
Users, whether enterprises or individuals, may run downloadable (i.e., open release) \citep{solaiman_gradient_2023} agents either on cloud compute or on their own hardware. A user may even be able to combine systems from different deployers to form an agent. 
Although visibility on the resulting agent may be desirable, the individual systems may not be significant enough by themselves to justify implementation of visibility measures by deployers. 
Indeed, a malicious actor could build and run an agent in this way so as to avoid detection by regulatory authorities. In this section, we discuss how our visibility measures may be extended to such situations, as well as the risks of doing so.

\subsection{Compute Provider Oversight}
Compute providers could enable oversight over deployments that involve large quantities of compute. Large-scale deployments could be concerning because they might involve vast numbers of agents, which could translate into a large impact multiplier for the user. Large-scale deployments are also noticeable because they consume significant resources. Compute providers may have cost advantages over users deploying their own hardware because of economies of scale. Indeed, using compute as a service (e.g., infrastructure as a service or cloud) is the default way for a business to deploy its IT services. If a compute provider can identify large-scale deployments and whether they correspond to agent activities, they may ask the user for proof that they have implemented certain visibility measures \citep{egan_oversight_2023,obrien_deployment_2023}. 

\subsection{Tool and Service Providers as Distributed Enforcement Mechanisms}
The need for agents to interact with external tools offers another leverage point. By conditioning tool and service access on implementation of certain visibility measures such as agent identifiers, tool and service providers can incentivize adherence to the measures. For example, financial institutions could restrict access to AI agents without identifiers from certain trusted deployers. Such identifiers might explicitly confirm permissions to access certain services, such as performing financial transactions or accessing certain websites. This approach could also allow tool providers to minimize misuse and enable detailed analytics of AI agent interactions with their tools.

One limitation is that AI agents could circumvent APIs by directly interacting with tools in a way that mimics human behavior. 
The development of tools capable of detecting disguised AI activity, akin to modern CAPTCHA systems designed to differentiate between human and software interactions, may be helpful. 
An alternative is to require proof of human identity for high-risk actions. Certain industries perform identity verification for high-risk activities with know-your-customer protocols \citep{egan_oversight_2023}. Similarly, tool or service providers operating in high-risk domains could require human identification. 
One difficulty is preventing AI agents from spoofing humans, such as through generating fake identification documents or stealing real ones. While CAPTCHA-like tests are a possibility, measures should be robust to improvements in the capabilities of agents. 
How to balance privacy considerations with the need for identity verification is another open question. A potential direction is to understand what mechanisms may allow humans to prove their status without identity disclosure. 

While direct interaction with tools is possible, users and developers may still opt for the convenience and efficiency offered by APIs, especially if direct interaction is more complex. 
APIs can provide standardized interfaces, tailored services for AI agent use, and can set specific conditions like access rates or the scope of services available. 
This preference for API interaction could reduce the difficulty of obtaining visibility into decentralized deployments. 

\subsection{Risks}
Extending visibility measures to decentralized deployments has serious implications for privacy and concentration of power. Compute providers that surveil deployments may be able to infer sensitive information about users. Given that a handful of compute providers dominates the market \citep{richter_infographic_2023}, monitoring users of those providers would equivalent to monitoring much of society. 
In addition to potential abuses of collected information, compute providers may also have lax security standards that enable attackers to gain sensitive information.  

Enforcement through tool and service providers also faces similar concerns. If useful tools and services were unavaiable to agents that were not from certified deployers, users may face strong pressure to use agents from such deployers. Whether because of government demand \citep{greenwald_nsa_2013} or regulatory capture \citep{dal_bo_regulatory_2006}, those deployers may have practises that are inimical to users or may not be responsive to user interests. Visibility measures for those deployers may be extremely invasive, such as excessive and unjustified logging. Agents from those deployers may not be well-suited to the user's use cases; for example, the user might require an agent to be able to operate in a low-resource language. If the market of deployers is heavily concentrated, further reliance upon them could exacerbate systemic risks \citep{widder_open_2023}. 

One way to mitigate these risks is to explore voluntary standards for adopting visibility measures. Voluntary standards could allow experimentation to understand when and where visibility measures should be applied. Although voluntary standards might not provide visibility into malicious use or enjoy universal adherence, understanding gained from their adoption can aid their later codification.

Certain tools may aid the adoption of voluntary standards. For example, open-source frameworks to implement agent identifiers may allow users to avoid deployers and, at the same time, facilitate visibility. Even if tool and service providers reject requests from agents without identifiers, users may easily be able to add an identifier to their agents. Independent entities may be required to certify valid identifiers, similar to certificate authorities on the Internet. A source of inspiration may be Let's Encrypt, a non-profit certificate authority which provides a free, automatic certificate process and which has been instrumental in promoting the use of the more secure HTTPS \citep{lets_encrypt_lets_2024}. Financial and technical support to develop open-source, agent identifier frameworks will likely be critical to the success of voluntary standards.

Another potential mitigation is to limit the number of tools or services that require agent identifiers. Obtaining visibility over tools involved in potentially high-risk activities, such as scientific platforms that handle pathogens or dangerous chemicals \citep{boiko_emergent_2023}, may be of higher priority. Similarly, visibility into business uses of AI agents, rather than personal uses, may be more important. Regulations could require certain businesses (e.g., those above a certain size) that use AI agents to implement certain visibility measures. 

Rather than mandating denial of requests from agents that do not have identifiers or that do not provide proof that certain visibility measures are implemented, one alternative is accounting for such compliance when determining the legal liability of users who deploy their own agents. Analogously, compliance with HIPAA de-identification standards can be taken into consideration to reduce regulatory fines or audits for violations \citep{noauthor_42_nodate,us_department_of_human_and_health_services_guidance_2022}. Similarly, the 2023 U.S. National Cybersecurity Strategy proposes to shield from private liability companies that follow cybersecurity best practices \citep{the_white_house_national_2023}. Accounting for compliance when determining liability may incentivize standards adherence. 

Other potential mitigations to explore include decentralized data custody schemes for compute provider logs and enhanced transparency into both data collection practises and government requests for data \citep{goodwin_cooperation_2018}.

\section{Conclusion}
Visibility facilitates the governance of increasingly agentic systems. We assessed three mechanisms for visibility: agent identifiers, real-time monitoring, and activity logs. 
Agent identifiers indicate whether and which agents are involved in an interaction. To aid accountability and incident investigation, an agent card containing additional information about the agent can be attached to an agent identifier. 
Real-time monitoring aims to flag problematic agent behaviour as it happens. Activity logs record certain inputs and outputs of agents so as to enable in-depth, post-hoc analysis of behaviour. 
We examined how to extend the visibility measures to decentralized deployments of agents, in particular through using compute providers and tools and services providers to obtain visibility. 
Finally, we analyzed the implications of the visibility measures on privacy and concentration of power. Rather than advocating for immediate implementation of these measures, further understanding of the measures and how to mitigate their negative impacts is required. 
Such understanding can help to build a foundation for the governance of AI agents.

Visibility informs actions to manage risks from the deployment of increasingly agentic systems, but is not by itself sufficient. Even with a comprehensive understanding of agent activities, those harmed by them may not have the power to intervene and reduce risks \citep{ananny_seeing_2018}. 
To best make use of visibility, future work could investigate increasing public influence over AI development and deployment \citep{chan_reclaiming_2023,huang_generative_2023,ovadya_generative_2023,seger_democratising_2023}, developing a wide range of potential policy levers \citep{anderljung_frontier_2023,egan_oversight_2023,shavit_practices_2023}, and implementing infrastructure and practices to prevent or defend against harms \citep{buterin_my_2023,sandbrink_differential_2022}.

\begin{acks}
We are grateful to the following people for insightful feedback and conversations over the course of writing this work: Charlotte Siegmann, Chinmay Deshpande, Nathan Barnard, Leonie Koessler, Kevin Frazier, Matthijs Maas, Micah Carroll, Yonadav Shavit, Merlin Stein, Ben Bucknall, Shalaleh Rismani, Paul Crowley, Gabriel Recchia, Janet Egan.
\end{acks}

\bibliographystyle{ACM-Reference-Format}


\pagebreak

\appendix

\section{Potential Information to Include on an Agent Card}\label{app:agent-card}
\subsection{The Underlying System}
\begin{itemize}
    \item Evaluations of the system's degree of agency \citep{mialon_gaia_2023,ruan_identifying_2023,liu_agentbench_2023};
    \item Evaluations of the system's generality: its ability to accomplish a broad array of tasks to some specified performance threshold \citep{morris_levels_2023};
    \item Red flags, such as previous incidents or results of previous dangerous capabilities and alignment evaluations \citep{shevlane_model_2023}; 
    \item Dependencies of the agent, such as with an ecosystem graph \citep{bommasani_ecosystem_2023} (e.g., whether the agent is a fine-tuned variant of another model).
\end{itemize}

\subsection{The Specific Instance of the Agent}
\begin{itemize}
    \item  How the agent instance was created (e.g., by its user directly or by another agent?);
    \item  The agent's goal, including both what the user specifies and what the system appears to be achieving \citep{gilbert_reward_2023};
    \item  Any tools or services that the agent can access (e.g., spinning up another agent through an API call, physical manipulation of robotics);
    \item  The agent's permissions (e.g., whether it has sudo access in a terminal);
    \item  Details about the persistence of the agent, such as whether the agent has a set lifetime;
    \item  The sector of the intended deployment environment (e.g., finance);
    \item The number of people the system can directly impact and the severity of such impact;
    \item The degree and ease of human oversight over the agent.
\end{itemize}

\subsection{The Actors Involved in the Creation and Operation of the Agent}
\begin{itemize}
    \item The user;
    \item The compute provider;
    \item Developers of the underlying system, including developers of any scaffolding or component foundation models;
    \item Any humans with whom the agent has interacted;
    \item Tools and services providers that the agent can use.
\end{itemize}

\end{document}